# Polarimetric imaging with the GMRT

**Sanhita Joshi***
*National Centre for Radio Astrophysics*
*Pune University Campus*
*Pune, 411 007, India*
*E-mail:* sjoshi@ncra.tifr.res.in

**Jayaram Chengalur**
*National Centre for Radio Astrophysics*
*Pune University Campus*
*Pune, 411 007, India*
*E-mail:* chengalu@ncra.tifr.res.in

We present the first set of polarimetric images made with the GMRT. These were obtained as part of the program to commission the polarization mode at the telescope. We find that the instrumental polarization leakage at the GMRT varies with frequency. It is hence necessary to solve for the leakage as a function of spectral channel. Once this is done however, it is possible to calibrate these terms to better than 1% accuracy, making it feasible to study sources that are polarized at the few percent level. We present 610 MHz polarization images of two extended FR-II radio galaxies, viz. 3C 79 and 3C 265. These were selected from the sample of sources for which the total polarization fraction at 610 MHz is known from the survey of Conway & Strom (1984). We present high resolution polarization images of these two sources and also find that the polarization fractions of the two sources as seen at the GMRT are consistent with those reported by Conway & Strom (1984).

*Speaker.

## 1. Introduction

Although the magnetic fields in the intra-cluster medium may not be dynamically important they do play a major role in determining the heat conduction and transport of charged particles [3,8]. Radio synchrotron radiation, its polarization and Faraday rotation are important tools in studying magnetic fields, see [1] for a recent review. The strength of the magnetic field governs the total intensity of the radio synchrotron emission, whereas the polarized emission tells us about the direction of the magnetic fields and Faraday rotation is an indicator of the large scale ordered fields.

Here, we present GMRT 610 MHz polarization observations of two FR-II radio galaxies, viz. 3C 79 and 3C 265 that were obtained as part of the testing of the polarization mode at the GMRT.

## 2. Observations and Data reduction

Although the GMRT correlator has a polarization mode [3] there has, to date, been a limited amount of polarization work done in the interferometric mode with the GMRT. We present here data taken as part of the testing of the polarization mode and polarization leakage at the GMRT.

Fig. 1[A] shows the Stokes RL normalized to the total intensity as a function of channel number for the unpolarized source 3C 147. In all the observations, the observing bandwidth of 16 MHz was divided in to 128 frequency channels, each with 125 kHz width. All the data reduction was done using the NRAO Astronomical Image Processing Software (AIPS). The red and the blue curves are for two consecutive half an hour time slices respectively. As can be seen, the leakages vary across the band, and there is a slight variation in the leakage with time. Leakage calibration hence needs to be done separately per channel.

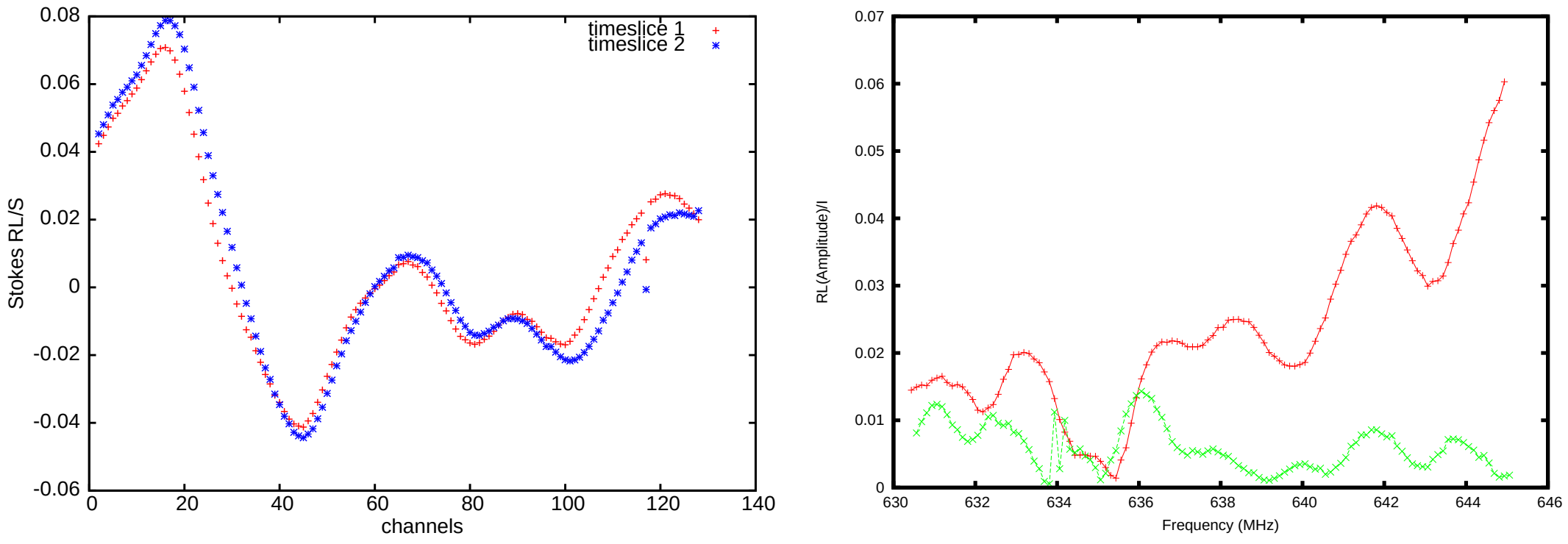

**Figure 1:** [A]On the left is the plot of Stokes RL with the total intensity for 3C 147, as a function of channels. The two curves are for two timeslices of half an hour each. [B] The right panel shows the Stokes RL normalized with Stokes I for the unpolarized calibrator 0318+164 before (red) and after (green) polarization calibration. The calibration was done using the leakages computed from the 3C 147 scans

For the sources 3C 279 and 3C 265 the observations were made at two different frequency settings, viz. 595 and 630 MHz. Scans at the two different settings were interleaved. The on source time was $\sim$ 4 hrs in each frequency. During both the observations, we observed 3C 147 as the flux, bandpass and polarization leakage calibrator, 0318+164 and 1227+365 as the phase calibrators for the target sources 3C 79 and 3C 265 respectively. The polarized calibrator 3C 286 was also in order to determine the polarization position angle during the 3C 265 observations. Fig. 1[B] shows Stokes RL normalized by stokes I for before and after calibration for the phase calibrator 0318+164. The calibration was done using the leakages measured for 3C 147. As can be seen, that after polarization calibration, the residual leakage is $\sim$ 1% or less per channel.

In the case of 3C 79, the data from only one of the two frequency settings viz. 630 MHz was found to be usable. The delay difference between the R and the L channels were calibrated using scans on the calibrator 3C 147. For 3C 265, the observations of the polarized calibrator 3C 286, were used to calibrate the position angle for each individual channels using the tasks RLDIF and CLCOR. While using the task CLCOR the phases for 3C 286 were also corrected for its rotation measure 1.2 rad-m$^{-2}$ [11]. No absolute polarization position angle correction is done for these data.

For the polarized targets, Stokes I data for 100 good channels were averages together to make Stokes I images. The total intensity map produced with a few iterations of imaging and self calibration, was used to calibrate the individual frequency channels. These data were used for Stokes Q, U and V mapping. For individual channels, the rms on these maps is $\sim$ 1mJy.

## 3. Results

Fig. 2[A] shows the distribution of the total polarized flux in 3C 79. The channels were averaged together using rotation measure synthesis [2]. Since the variation of position angle with channel was not calibrated for 3C 79, the rotation measure of the source cannot be disentangled from any systematic instrumental variation of position angle with frequency, however the total polarized flux should be reliably measured. 3C 79 has a very weak core at low frequencies. The western hotspot is the brightest feature of the map, it also is slightly non-collinear with the rest of the extended emission and the eastern hotspot. As can be seen most of the polarization in the source comes from the eastern hotspot. The polarization percentages measured at the GMRT is given in Tab. 1 along with those measured at the WSRT by [4]. The same source has been observed at higher wavelegnths, 1.4 and 4.9 GHz [10] and 8.4 GHz [6]. The distribution of the polarized flux we observe at 50-cm is consistent with the those observations. The extended emission of the western lobe and the core are Faraday depolarized at 1.4 GHz compared with that at 4.9 GHz [10]; we also see the same trend at 610 MHz band. The eastern hotspot is depolarized mainly due to the beam depolarization. For the same beam size, as at 610 MHz, this hotspot is almost depolarized even at 8.4 GHz [6].

Like 3C 79, 3C 265 is also an FR-II radio galaxy with weak core emission in the 610 MHz band. Also, both the sources show asymmetry in terms of the total intensity and polarized flux density of the two lobes. In the case of 3C 265, the Eastern hotspot is the brightest of two and also has most of the polarized flux. At 610 MHz, we detect extended emission towards west, which is barely detected at 1.4 GHz by [5] with similar resolution as at 610 MHz; this emission is not detected at 4.85 GHz or 8.4 GHz [7]. The extended emission is found to be weakly polarized at 1.4 GHz [5], however, we do not detect any polarization at 610 MHz. Similarly, although [5] find the hot spots in the western lobe to be polarized at high frequency (8.4 GHz) and high resolution (0.35 arcsec), we do not detect much polarization at 610 MHz.

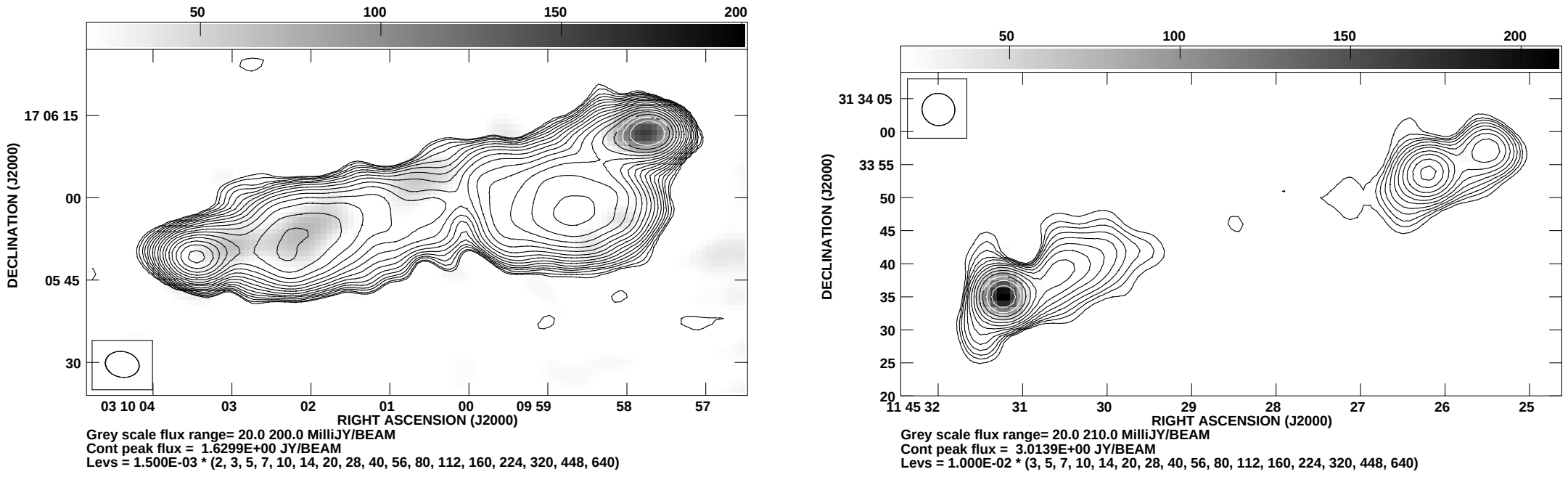

**Figure 2:** [A] The top panel shows 3C 79 (on the left) with contours showing the total flux density for 3C 79 averaged from the data at 630 MHz; the grey-scale shows the polarized flux density. The plot on the right is the same for the other source 3C 265, averaged for 595 and 630 MHz. The polarized flux density contours are drawn at about 5-$\sigma$ level, where $\sigma$ is the RMS noise near the source, measured in AIPS using TVSTAT. Resolution for 3C 79 is 8.4X5.5 arcmin and that for 3C 265 is 5.0X4.8 arcmin. The contours are drawn at levels 3mJy $\times$ 3,5,7,10,14,20,28,40,56,80

| Source | P (mJy) | I (Jy) | P (%) | I (Jy) * | P(%)* |
|---|---|---|---|---|---|
| 3C 79 | | | | | |
| Eastern lobe | 168 | 4.2 | 4.0 | 4.90 | 4.3 |
| Western lobe | 133 | 6.7 | 2.0 | 6.72 | 2.4 |
| 3C 265 | | | | | |
| Eastern lobe | 238 | 4.85 | 4.9 | 4.93 | 5.3 |
| Western lobe | 30 | 2.65 | 1.1 | 2.70 | 1.5 |

**Table 1:** Polarization fractions for the E and W lobes of the two sources. The last two columns are from [4].